\documentclass[twocolumn,aps,showpacs,floatfix,superscriptaddress]{revtex4}
\pdfoutput=1 
\usepackage{amsmath,amssymb,eucal,graphicx}
\begin{document}

\title{Reaction-Diffusion Process Driven by a Localized Source:
First Passage Properties}
\author{P. L. Krapivsky}
\affiliation{Department of Physics, Boston University, Boston, MA 02215, USA}
\begin{abstract}
We study a reaction-diffusion process that involves two species of atoms, immobile and diffusing. We start with only immobile atoms uniformly distributed throughout the entire space. Diffusing atoms are injected at the origin by a source which is turned on at time $t=0$. When a diffusing atom collides with an immobile atom, the two atoms form an immobile stable molecule. The region occupied by molecules is asymptotically spherical with radius growing as $t^{1/d}$ in $d\geq 2$ dimensions. We investigate the survival probability that a diffusing atom has not become a part of a molecule during the time interval $t$ after its injection. We show that asymptotically the survival probability (i) saturates in one dimension, (ii) vanishes algebraically with time in two dimensions (with exponent being a function of the dimensionless flux and determined as a zero of a confluent hypergeometric function), and (iii) exhibits a stretched exponential decay in three dimensions. 
\end{abstract}
\pacs{05.40.-a, 82.20.-w, 81.65.-b}
\maketitle

\section{Introduction}

Diffusion is the basic transport mechanism that underlies numerous phenomena in physics, chemistry, and biology \cite{hcb,w,rg,book}. In applications diffusing particles usually interact with each other, or with other species of particles. Here we consider a system in which diffusing atoms (species $A$) are injected into a small localized region of a $d-$dimensional lattice. The entire lattice is initially occupied by immobile atoms (species $B$), one $B$ atom per lattice site. Whenever an $A$ atom hops to a lattice site occupied by a $B$ atom, two atoms react to form an inert stable molecule (species $B^*$). Thus the process is described by the reaction scheme  \cite{Meakin,Lar}
\begin{equation*}
A(\text{diffusing})+B(\text{substrate})\rightarrow B^*(\text{stable})
\end{equation*}
This model and its generalizations mimic important industrial chemical processes such as electropolishing \cite{EP}, dissolution of solids \cite{D87}, corrosion and etching \cite{Meakin,Lar,KM91}, and erosion \cite{SBG}. In the injection-controlled limit, namely in the situation when injection is so slow that each new $A$ atom will undergo reaction before the next injection event, the process is also known as internal diffusion limited aggregation (IDLA). 

The reaction $A+B\rightarrow B^*$ proceeds at a certain rate which often greatly exceeds the hopping rate. Therefore we shall always assume that the reaction proceeds instantaneously, so each site contains either a $B$ atom or a $B^*$ molecule. Hence as the process develops, the system can be separated onto two parts: The droplet that contains no $B$ atoms (every lattice site inside the droplet is occupied by a $B^*$ molecule and can contain an arbitrary number of $A$ atoms) and the rest of the system that contains only $B$ atoms. 

The droplet is of course a growing random set, but in the large time limit it becomes relatively more and more close to the ball. (This is not entirely obvious, e.g. external diffusion limited aggregates \cite{WS} are not spherical at all, their shape depend on the lattice and they have very intricate branching structures.) For IDLA, however, the assertion that the asymptotic shape of the droplet is a ball has been proved \cite{BGL}; the general case when the strength of the source is finite has been also settled \cite{Stefan}. Intuitively, one expects \cite{Lar} that in the long time limit the radius of the droplet $R(t)$ and the density of $A$ atoms inside the droplet, can be found by solving a Stefan-like problem, namely by solving a diffusion equation with a moving boundary whose position is determined in the process of solution \cite{CJ59,C87}. Gravner and Quastel proved \cite{Stefan} that this is indeed correct. Using this reduction to the Stefan problem one finds \cite{Lar,Stefan} that the radius of the droplet scales as
\begin{equation}
\label{radius}
R\propto
\begin{cases}
\sqrt{t\ln t}   &d=1\\
t^{1/d}         &d\geq 2
\end{cases}
\end{equation}
when $t\to\infty$. 

Earlier simulations \cite{Meakin,KM91} revealed that the surface of the droplets is strikingly smooth, e.g. in two dimensions the width of the interface appears to scale logarithmically with droplet radius. This logarithmic growth law has resisted the proof up until recently when it was shown (in the IDLA setting) that the width grows not faster than logarithmically \cite{IDLA_0,IDLA_1,IDLA_2,IDLA_3} with the size of the droplet, more precisely not faster than $\ln R$ when $d=2$ \cite{IDLA_1,IDLA_2} and not faster than $\sqrt{\ln R}$ when $d\geq 3$ \cite{IDLA_1,IDLA_3}. 

In this article we examine first passage properties of diffusing atoms. One particularly interesting quantity is the survival probability $S(t_0,t)$ that an $A$ atom injected at a certain time $t_0$ has not reached the (growing) outer boundary during the time interval $(t_0,t)$.  We shall mostly focus on the limit when the observation time greatly exceeds the release time: $t\gg t_0$. We shall show that in this limit the surviaval probability varies with the observation time $t$ according to 
\begin{equation}
\label{surv_asymp}
S\propto
\begin{cases}
\text{finite}       &d=1\\
t^{-\beta}         &d=2\\
\exp\!\left[-\text{const}\times t^{1/3}\right]         &d=3
\end{cases}
\end{equation}
where the exponent $\beta$ depends on the (dimensionless) ratio of the flux rate to the hopping rate. 

An outline of this paper is as follows. In Sect.~\ref{Stefan-problem}, we review the general formulation of the Stefan problem and the solution of the Stefan problem in two dimensions. The two-dimensional case is most common in applications as a localized injection from the `third' dimension can naturally occur. The two-dimensional case is also mathematically more appealing as the Stefan problem admits a self-similar solution. In Sect.~\ref{Survival_2} we study the behavior of the survival probability in the two-dimensional setting. The analysis in the three-dimensional case is presented in Sect.~\ref{Stefan_3}. In Sect.~\ref{convection} we returned to two dimensions, but assume that mobile $A$ atoms are injected with a fluid, so their transport is determined both by convection and diffusion. Finally in Sect.~\ref{summary} we give a brief summary.

\section{Stefan Problem in Two Dimensions}
\label{Stefan-problem}

Consider a $d-$dimensional lattice initially occupied by immobile atoms (species $B$). Let diffusing atoms (species $A$) be injected at the origin. Whenever a mobile $A$ atom lands on a lattice site with an immobile $B$ atom, the two atoms immediately form a $B^*$ molecule:
\begin{equation}
\label{scheme}
A+B\rightarrow B^*
\end{equation}
The molecules $B^*$ are assumed to be immobile and stable. The source of $A$ atoms is turned on at time $t=0$. The droplet that contains no $B$ atoms is asymptotically a ball of radius $R(t)$ and its boundary moves more and more deterministically when $t\gg 1$. In the hydrodynamic framework, the concentration $c({\bf r},t)$ of $A$ atoms (that is, the average number of $A$ atoms per lattice site) satisfies a diffusion equation 
\begin{equation}
\label{RD:c2d}
\frac{\partial c(r,t)}{\partial t} = D\!\left(\frac{\partial^2}{\partial r^2} + 
\frac{d-1}{r}\,\frac{\partial}{\partial r}\right)\!c(r,t) + J\delta({\bf r})\,\theta(t)
\end{equation}
inside the droplet $0\leq r\leq R(t)$. Here $D$ is the hopping rate of $A$ atoms, $J$ is the strength of the source, and $\theta(t)$ is the Heaviside step function. 

The diffusion equation \eqref{RD:c2d} should be supplemented by the standard absorbing condition
\begin{equation}
\label{BC:c}
c(r=R(t), t>0)=0
\end{equation}
and the Stefan boundary condition
\begin{equation}
\label{Stefan}
\frac{dR}{dt} = - D\frac{\partial c}{\partial r}\Big|_{r=R} 
\end{equation}
which essentially expresses mass conservation (see \cite{CJ59,C87} for derivations of such boundary conditions in various examples). The boundary moves and its position has to be determined in the process of solution. Therefore mathematically we arrive at the Stefan problem \cite{CJ59,C87}. 

The original process occurs on the lattice and therefore we set the lattice spacing to unity; this implies that the spatial coordinates ${\bf r}$, the droplet radius $R(t)$, and the concentration $c({\bf r},t)$ are all dimensionless quantities, while the hopping rate $D$ and the strength of the source have the dimension of inverse time: $[D]=[J]=1/(\text{time})$. Note that the ratio $J/D$ is dimensionless and this parameter plays an important role in the problem. It will prove convenient to use a slightly modified ratio
\begin{equation}
\label{flux}
\Phi = \frac{J}{4\pi D}
\end{equation}
which we shall call the dimensionless flux. 

We now focus on the two-dimensional case; we shall discuss the three-dimensional set-up in Sect.~\ref{Stefan_3}. In two dimensions, the Stefan problem \eqref{RD:c2d}--\eqref{Stefan} has been solved in \cite{Lar}. Here we outline some details of the solution as we shall need them for the analysis of the survival probability. One seeks solution in the scaling form
\begin{equation}
\label{scaling}
c(r,t) = c(\xi), \quad \xi=\frac{r}{R}
\end{equation}
The derivatives of the density read
\begin{equation}
\label{deriv}
\frac{\partial c}{\partial t} = -\frac{\dot R}{R}\,\xi c',
\quad \frac{\partial c}{\partial r} = \frac{1}{R}\,c', 
\quad \frac{\partial^2 c}{\partial r^2} = \frac{1}{R^2}\,c''
\end{equation}
where $(\cdot)'\equiv d(\cdot)/d\xi$ and $\dot R=dR/dt$. The governing diffusion equation \eqref{RD:c2d} reduces to
\begin{equation}
\label{RD:c2d-scal}
c'' + \frac{1}{\xi}\,c' = - \frac{R\dot R}{D}\,\xi c'
\end{equation}
This equation is consistent if $R\dot R/D$, which is in principle a function of time, is actually a constant. Denoting this (yet unknown) constant by $2\alpha$ we get
\begin{equation}
\label{radius:2d}
R^2 = 4\alpha Dt
\end{equation}
and recast \eqref{RD:c2d-scal} into 
\begin{equation}
\label{c2d-scal}
c'' + (\xi^{-1}+2\alpha \xi)\,c' = 0
\end{equation}
A solution to Eq.~\eqref{c2d-scal} satisfying the adsorbing boundary condition \eqref{BC:c} reads
\begin{equation}
\label{c:2d}
c(\xi)=A\int_\xi^1 \frac{d\eta}{\eta}\,e^{-\alpha\eta^2}
\end{equation}
Plugging this solution to the Stefan boundary condition \eqref{Stefan} allows one to express the amplitude $A$ via $\alpha$:
\begin{equation}
\label{Aa}
A = 2\alpha\,e^{\alpha}
\end{equation}
To determine $\alpha$, we notice that 
\begin{equation}
\label{conserv_law:2d}
Jt = \int_0^R c(r,t)\,2\pi r\,dr + \pi R^2
\end{equation}
Equation \eqref{conserv_law:2d} is just the conservation of the total number of $A$ atoms. Re-writing 
\eqref{conserv_law:2d} through the scaling variables \eqref{scaling} and using
\eqref{radius:2d} we arrive at relation 
\begin{equation}
\label{a:JD}
\alpha\,e^{\alpha} = \Phi
\end{equation}
which implicitly determines $\alpha$ in terms of the dimensionless flux $\Phi$. Note that the fraction of diffusing $A$ atoms (that is, those which are not the part of $B^*$ molecules) with respect to all emitted $A$ atoms is $1-e^{-\alpha}$. Put differently, the total number $\mathcal{A}(t)$ of diffusing $A$ atoms grows linearly with time: $\mathcal{A}(t)\simeq (1-e^{-\alpha})Jt$.

\section{Survival Probability in Two Dimensions}
\label{Survival_2}

Although the Stefan problem described in Sect.~\ref{Stefan-problem} has been previously understood as far as the general growth characteristics are concerned, more subtle features characterizing the behavior of diffusing atoms haven't been explored. Here we analyze first passage properties \cite{fpp} of diffusing atoms. As an example, consider an $A$ atom that has been injected at a certain time $t_0$ and ask for the survival probability $S(t_0,t)$, namely the probability that it has not reached the (growing) outer boundary during the time interval $(t_0,t)$. The two-variable survival probability $S(t_0,t)$ apparently has an interesting scaling form in the limit when both $t_0$ and $t$ are very large and comparable, that is, 
\begin{equation}
\label{double_scaling}
t_0\to\infty, \quad t\to\infty, \quad \frac{t_0}{t}=\text{finite} 
\end{equation}

More simple behavior should arise when the observation time greatly exceeds the release time, that is, $t\gg t_0$. Indeed, the average distance from the origin exhibits the diffusive growth, $\langle r^2\rangle=4D(t-t_0)$, which is similar to the growth of the radius of the droplet, $R=\sqrt{4D\alpha t} $. This suggests that the survival probability exhibits an algebraic decay
\begin{equation}
\label{survival:2d}
S(t_0,t) \propto t^{-\beta} \quad \text{when}\quad t\gg t_0
\end{equation}
The persistence exponent $\beta$ should be a function of $\alpha$. The dependence $\beta=\beta(\alpha)$ can indeed be analytically determined. We shall employ essentially the same method as in Refs.~\cite{book,fpp,RK,FM}. 

To determine the survival probability (in the $t\gg t_0$ limit) we look at a more detailed quantity, the probability distribution $P(r,t)$. By definition, $P({\bf r},t)\,d{\bf r}$ is the probability that the atom has never reached the growing outer boundary of the droplet during the time interval $(t_0,t)$ and at the final time the atom lies in the region of area $d{\bf r}$ around ${\bf r}$. One anticipates that the probability distribution $P(r,t)$ is actually a function of the scaled spatial coordinate $\xi=r/R(t)$. Therefore we make a scaling ansatz similar to \eqref{scaling}
\begin{equation}
\label{scaling_P}
P(r,t) =  \frac{S(t)}{2\pi R^2(t)}\,\mathcal{P}(\xi), \quad \xi=\frac{r}{R(t)}
\end{equation}
We shortly write $S(t)$ instead of $S(t_0,t)$ since we assume that $t_0\ll t$ and we are mostly interested in the dependence on $t$. The time-dependent pre-factor in \eqref{scaling_P} has been chosen to ensure the validity of the connection between the probability distribution $P(r,t)$ and the survival probability. Indeed, 
\begin{eqnarray*}
S(t) = \int_0^R P(r,t)\, 2\pi r\,dr = S(t)\int_0^1d\xi\,\xi\, \mathcal{P}(\xi)
\end{eqnarray*}
and therefore the consistency condition is 
\begin{equation}
\label{P:condition}
\int_0^1 d\xi\,\xi \mathcal{P}(\xi) = 1
\end{equation}
The probability distribution $P(r,t)$ satisfies the diffusion equation 
\begin{equation}
\label{P:2d}
\frac{\partial P}{\partial t} = D\left(\frac{\partial^2}{\partial r^2} + 
\frac{1}{r}\,\frac{\partial}{\partial r}\right)P
\end{equation}
in the region $0\leq r\leq R(t)$. There is no source, apart from the fact that initially the atom is at the origin.  Hence the initial condition is 
\begin{equation}
\label{IC:P}
P({\bf r}, t_0) = \delta({\bf r})
\end{equation}
and the adsorbing boundary condition is
\begin{equation}
\label{BC:P}
P(r=R(t'), t')=0
\end{equation}
for $t_0<t'<t$. In the $t\gg t_0$ limit, the survival probability exhibits a power-law dependence $S(t_0,t) \propto t^{-\beta}$ on the observation time, and as long as we are interested in the persistence exponent $\beta$ we can forget on $t_0$. 

Using \eqref{scaling_P} and \eqref{radius:2d} we recast the diffusion equation \eqref{P:2d} into an ordinary differential equation 
\begin{equation}
\label{P2d-scal}
\mathcal{P}'' + (\xi^{-1}+2\alpha \xi)\,\mathcal{P}' +4\alpha(1+\beta)\mathcal{P}= 0
\end{equation}
The change of the variable $\xi$ to 
\begin{equation}
\label{zeta:def}
\zeta = -\alpha\xi^2
\end{equation}
transforms \eqref{P2d-scal} into a hypergeometric equation 
\begin{equation}
\label{P2d-zeta}
\zeta\,\frac{d^2 \mathcal{P}}{d\zeta^2} + (1-\zeta)\,\frac{d \mathcal{P}}{d\zeta} 
-(1+\beta)\mathcal{P}= 0
\end{equation}
The solution to \eqref{P2d-zeta} must satisfy the adsorbing boundary condition \eqref{BC:P}, or equivalently
\begin{equation}
\label{P:BC}
\mathcal{P}(\xi=1)=0,
\end{equation}
and the integral requirement \eqref{P:condition}. It should also be regular at $r=0$.

The condition of regularity at $r=0$ selects a solution up to an amplitude, 
$\mathcal{P} = C\,F(1+\beta, 1; \zeta)$, where $F$ is the confluent hypergeometric function \cite{NIST}. Therefore 
\begin{equation}
\label{Pscal:solution}
\mathcal{P} = C\,F(1+\beta, 1;  -\alpha\xi^2)
\end{equation}
and the boundary condition \eqref{P:BC} yields an equation 
\begin{equation}
\label{ba:implicit}
F(1+\beta, 1;  -\alpha)=0
\end{equation}
which relates the persistence exponent $\beta$ and the growth constant $\alpha$. 

The $\beta=\beta(\alpha)$ dependence provided by Eq.~\eqref{ba:implicit} is implicit. This equation has many solutions, the proper one corresponds to minimal $\beta$.  Having determined $\beta=\beta(\alpha)$, we can fix the amplitude $C$ in Eq.~\eqref{Pscal:solution} by requiring the validity of \eqref{P:condition}. The final result reads
\begin{equation}
\label{Pscal:sol}
\mathcal{P} = 2\,\frac{F(1+\beta, 1;  -\alpha\xi^2)}{F(1+\beta, 2;  -\alpha)}
\end{equation}

More explicit results can be established in certain special cases.  For instance, since 
$\beta\to \infty$ as $\alpha\to 0$, we can asymptotically re-write the confluent hypergeometric function $F\equiv F(1+\beta, 1;  -\alpha)$ in terms of the Bessel function of zero order:
\begin{eqnarray}
\label{Bessel_0}
F &=& 1 - \frac{(1+\beta)\alpha}{(1!)^2}+\frac{(1+\beta)(2+\beta)\alpha^2}{(2!)^2}-\ldots \nonumber\\
   &\simeq& 1- \frac{\beta\alpha}{(1!)^2}+ \frac{(\beta\alpha)^2}{(2!)^2}-\ldots \nonumber\\
   &=& J_0\big(2\sqrt{\alpha\beta}\big)
\end{eqnarray}
The Bessel function has infinitely many zeros, $J_0(x)=0$ when $x=\pm x_1,\pm x_2,\ldots$. The relevant root is closest to the origin, namely $x_1=2.404825558\ldots$. Therefore $\beta=(x_1/2)^2/\alpha=1.445796491/\alpha$ in the $\alpha\to 0$ limit.  When $\alpha$ is small, the dimensionless flux is also small, $\Phi\simeq \alpha$ according to Eq.~\eqref{a:JD}, and hence
\begin{equation}
\beta = \frac{1.445796491}{\Phi} = 18.16841454\,\frac{D}{J}
\end{equation}
when $D\gg J$. In this limit the probability distribution also acquires a neat form 
\begin{equation}
\label{P:limit}
\mathcal{P} = \frac{x_1 J_0(x_1\xi)}{J_1(x_1)}
\end{equation}

The confluent hypergeometric function simplifies when the indexes are integer. For instance, let $\beta=1$. Since $F(2,1;-\alpha)=(1-\alpha)e^{-\alpha}$, equation \eqref{ba:implicit} yields $\alpha=1$. Recalling \eqref{a:JD} we conclude that 
\begin{equation*}
S\sim t^{-1}\quad \text{when}\quad \Phi=e
\end{equation*}
and the probability density in this case is 
\begin{equation*}
\mathcal{P} = 2(1-\xi^2)\,e^{1-\xi^2}
\end{equation*}
For $\beta=2$, we have 
$F(3,1;-\alpha)=\big(1-2\alpha+\tfrac{1}{2}\alpha^2\big)e^{-\alpha}$. Relation \eqref{ba:implicit} leads to $1-2\alpha+\tfrac{1}{2}\alpha^2=0$, from which $\alpha=2-\sqrt{2}$ and therefore 
\begin{equation*}
S\sim t^{-2}\quad \text{when}\quad \Phi=\big(2-\sqrt{2}\big)e^{2-\sqrt{2}}
\end{equation*}

Previous scaling analysis does not allow one to probe the dependence of the two-variable survival probability $S(t_0,t)$ on the release time $t_0$. On dimensional grounds, one anticipates a simple algebraic behavior
\begin{equation}
\label{survival:2d_full}
S(t_0,t) \sim \left(\frac{t_0}{t}\right)^\beta 
\end{equation}
Equation \eqref{survival:2d_full} is expected to hold in the same $t\gg t_0$ limit as \eqref{survival:2d}; in contrast to the latter, however, Eq.~\eqref{survival:2d_full} has a proper dimensional form. Moreover, one anticipates that the survival probability remains finite in the double scaling limit \eqref{double_scaling} and Eq.~\eqref{survival:2d_full} agrees with this requirement. This suggests that Eq.~\eqref{survival:2d_full}  gives an approximately correct behavior in the entire time range.

\section{Three Dimensions}
\label{Stefan_3}

We begin again with the Stefan problem. In three dimensions, the governing diffusion equation reads
\begin{equation}
\label{RD:c3d}
\frac{\partial c}{\partial t} = D\left(\frac{\partial^2}{\partial r^2} + 
\frac{2}{r}\,\frac{\partial}{\partial r}\right)c 
\end{equation}
and we seek a solution to \eqref{RD:c3d} inside the growing droplet $0\leq r\leq R(t)$. The boundary conditions are the same as in two dimensions, Eqs.~\eqref{BC:c}--\eqref{Stefan}. Using again the scaling ansatz \eqref{scaling} we reduce \eqref{RD:c3d} to 
\begin{equation}
\label{RD:c3d-scal}
c'' + \frac{2}{\xi}\,c' = - \frac{R\dot R}{D}\,\xi c'
\end{equation}
The conservation law 
\begin{equation}
\label{conserv_law:3d}
Jt = \int_0^R c(r,t)\,4\pi r^2\,dr + \frac{4\pi}{3}\, R^3
\end{equation}
tells us that $Jt>\tfrac{4\pi}{3} R^3$. Hence the radius grows at most as
$t^{1/3}$ and $R\dot R$ decays at least as $t^{-1/3}$. Therefore the right-hand side of equation \eqref{RD:c3d-scal} asymptotically vanishes and we get $c'' + \frac{2}{\xi}\,c' = 0$ from which 
\begin{equation}
\label{c3d:sol}
c=C\left(\frac{1}{\xi}-1\right)
\end{equation}
Near the origin the density is stationary and it satisfies 
$\nabla^2 c = -\tfrac{J}{D}\delta({\bf r})$. Therefore $c=\Phi/r$ when $r\ll R$. Comparing this asymptotic with \eqref{c3d:sol} we find $C=\Phi/R$. This allows us to re-write \eqref{c3d:sol} as
\begin{equation}
\label{c3d:solution}
c=\Phi\left(\frac{1}{r}-\frac{1}{R}\right)
\end{equation}

Note that the number of diffusing $A$ atoms is
\begin{equation}
\label{free:3d}
\mathcal{A}(t) = \int_0^R c(r,t)\,4\pi r^2\,dr = \frac{J}{D}\,\frac{R^2}{6}
\end{equation}
The number of bounded $A$ atoms (which are included into $B^*$ molecules) scales as $R^3$ and therefore it constitutes the dominant part of all injected $A$ atoms. Hence $Jt\simeq \frac{4\pi}{3}R^3$ implying that asymptotically the radius of the droplet is given by
\begin{equation}
\label{radius:3d}
R = \left(\frac{3Jt}{4\pi}\right)^{1/3}
\end{equation}
Note that the fraction of injected $A$ atoms which continue to diffuse, $(Jt)^{-1}\mathcal{A}(t)$, decays as 
$\frac{1}{6}(3\Phi)^{2/3}(Dt)^{-1/3}$. 

Since the radius grows slower than diffusively, the survival probability should decay very quickly. To establish this decay qualitatively we must solve the diffusion equation 
\begin{equation}
\label{P:3d}
\frac{\partial P}{\partial t} = D\left(\frac{\partial^2}{\partial r^2} + 
\frac{2}{r}\,\frac{\partial}{\partial r}\right)P
\end{equation}
subject to the initial-boundary conditions \eqref{IC:P}--\eqref{BC:P}. The behavior is again simpler than in two dimensions. The leading asymptotic is
\begin{equation}
\label{3d:lowest}
P(r,t) = f(t)\,\frac{\sin(\pi r/R)}{r}
\end{equation}
where $r^{-1}\sin(\pi r/R)$ is the eigenfunction of the Laplace operator corresponding to the smallest eigenvalue. The dominance of the eigenfunction corresponding to the `ground state' is easy to appreciate --- since the growth is slow, $R\ll \sqrt{Dt}$, the density has enough time to equilibrate, so its spatial behavior is the same as in the case of the fixed-size droplet and only the time-dependent factor $f=f(t)$ may differ. For more details (in the one-dimensional setting) see \cite{RK}. 

Using \eqref{3d:lowest} we find
\begin{subequations}
\begin{align}
\label{time_P}
\frac{1}{P}\,\frac{\partial P}{\partial t} & =  \frac{\dot f}{f} 
- \frac{\pi r}{R}\,\frac{\dot R}{R}\\
\label{space_P}
\frac{1}{P}\,\nabla^2 P & = -\left(\frac{\pi}{R}\right)^2
\end{align}
\end{subequations}
It turns out that the first term on the right-hand side of \eqref{time_P} is dominant.
Hence $\dot f  = -D\big(\tfrac{\pi}{R}\big)^2 f$ from which
\begin{equation}
\label{general_f}
f(t)\sim \exp\!\left\{-D\int_0^t dt'\,\left[\frac{\pi}{R(t')}\right]^2\right\}
\end{equation}
Equation \eqref{general_f} is valid for arbitrary growth law of the radius which is slower than diffusive, $R\ll \sqrt{Dt}$, e.g. for any algebraic growth $R\sim t^a$ with $a<1/2$. Specializing the general result \eqref{general_f} to our case when the radius grows according to Eq.~\eqref{radius:3d} we obtain 
\begin{equation}
\label{special_f}
f\sim \exp\!\left\{-\pi^2\, \Phi^{-2/3}\, (3Dt)^{1/3}\right\}
\end{equation}
Note that $\dot f/f\sim t^{-2/3}$, while $\dot R/R\sim t^{-1}$. Hence the first term on the right-hand side of \eqref{time_P} is indeed dominant.

Finally, the survival probability is given by 
\begin{eqnarray*}
S &=& \int_0^R P(r,t)\,4\pi r^2\,dr\\
   &=& 4\pi f \left(\frac{\pi}{R}\right)^{-2}\int_0^\pi du\, u \sin u \qquad (u=\pi r/R)\\
   &=& 4fR^2
\end{eqnarray*}
The term $4R^2$ provides just a power-law correction to a controlling exponential decay of $f$. We have ignored such terms in a derivation of \eqref{general_f}--\eqref{special_f}. Therefore the survival probability decays according to the same law \eqref{special_f}, viz.
\begin{equation}
\label{survival:3d}
S\sim \exp\!\left\{-\pi^2\, \Phi^{-2/3}\, (3Dt)^{1/3}\right\}
\end{equation}
confirming the announced result \eqref{surv_asymp} in three dimensions. 

In the three-dimensional setting, one can analytically determine the asymptotic behavior of the two-variable survival probability $S(t_0,t)$ even when the realize time $t_0$ is {\em not} negligibly small in comparison with the observation time $t$. The probability density $P(0,t_0; r,t)$ describing this situation can be expressed using 
a quasi-static ansatz similar to \eqref{3d:lowest}, namely 
\begin{equation}
\label{3d:t_0}
P(0,t_0; r,t) = f(t_0,t)\,\frac{\sin(\pi r/R)}{r}
\end{equation}
This description is applicable when the typical diffusion length characterizing the released atom, $\sqrt{D(t-t_0)}$, greatly exceeds the droplet radius $R(t)$. Therefore using \eqref{radius:3d} we arrive at the criterion of the validity of \eqref{3d:t_0}
\begin{equation}
\label{3d:time_range}
t-t_0\gg D^{-1/3}\left(\frac{t}{\Phi}\right)^{2/3}
\end{equation}
Using \eqref{3d:t_0} instead of \eqref{3d:lowest} we find the controlling exponential behavior of the amplitude in Eq.~\eqref{3d:t_0}
\begin{equation}
\label{ft_0}
f(t_0,t)\sim \exp\!\left\{-D\int_{t_0}^t dt'\,\left[\frac{\pi}{R(t')}\right]^2\right\}
\end{equation}
which gives again the controlling exponential behavior of the two-variable survival probability
\begin{equation}
\label{survival:t_0}
S(t_0,t)\sim \exp\!\left\{-\frac{\pi^2}{\Phi^{2/3}} \left[(3Dt)^{1/3}-(3Dt_0)^{1/3}\right] \right\}
\end{equation}

It is tempting to use Eq.~\eqref{survival:t_0} to compute the number of diffusion $A$ atoms
\begin{equation}
\label{free:3d_sum}
\mathcal{A}(t) = J\int_0^t dt_0\, S(t_0,t)
\end{equation}
and to compare the outcome with the asymptotically exact prediction \eqref{free:3d}. The asymptotic results match if we use \eqref{survival:t_0} with numerical factor $\frac{1}{6}\pi^2$. This seemingly gives us the exact numerical amplitude and, more importantly, tells us that there is no power-law pre-factor to the prediction \eqref{survival:t_0}. A closer examination shows, however, that the main contribution to the integral on the right-hand side of Eq.~\eqref{free:3d_sum} is gathered in the region 
$t-t_0\sim D^{-1/3}(t/\Phi)^{2/3}$, which is precisely the temporal range [see \eqref{3d:time_range}] where the starting ansatz \eqref{3d:t_0} is no longer valid. Therefore we cannot use the sum rule \eqref{free:3d_sum} to fix the correcting pre-factors to the controlling exponential behavior, although it does seem plausible the power-law pre-factor is absent.

\section{Reaction-Convection-Diffusion Process with a Localized Source}
\label{convection}

In this section we investigate what happens if diffusion is supplemented by source-driven convection. The flux of diffusing $A$ atoms can be organized through the flux of the fluid containing $A$ atoms. Here we consider the most natural two-dimensional situation. We thus assume that  there is a flux of both $A$ atoms and fluid; as before, we posit that $B$'s and $B^*$'s are attached to the lattice and therefore immobile. This convection-diffusion problem may be interpreted as a pure diffusion problem \cite{KR} in a space with an effective dimension different from the physical dimension $d_{\rm phys}=2$. More precisely, let $Q$ be the fluid flux; put differently, the velocity field is given by $V=\tfrac{Q}{2\pi r}$. It turns out that this convection-diffusion problem can be recast as a pure diffusion that occurs in a space of effective dimension \cite{KR} 
\begin{equation}
\label{dim}
d=2-2q, \quad q=\frac{Q}{4\pi D}
\end{equation}
Therefore the analysis is similar to the one described in the previous sections. Most of the results which we will derive below hold for both sink and source flows, but the case of source flows ($Q>0$) is more natural, so we shall assume that we have a source of both $A$ atoms and fluid. Note that for source flows $q\geq 0$, so the effective dimension satisfies $d\leq 2$.

\subsection{Stefan Problem}
 
Generally in two spatial dimensions the convection-diffusion equation for radial velocity field $V=V(r)$ reads 
\begin{equation}
\frac{\partial c}{\partial t} + V\, \frac{\partial c}{\partial r}= D\left(\frac{\partial^2}{\partial r^2} + 
\frac{1}{r}\,\frac{\partial}{\partial r}\right)c 
\end{equation}
For the velocity field $V=\tfrac{Q}{2\pi r}$ this convection-diffusion equation becomes 
\begin{equation}
\label{conv_diff:2d}
\frac{\partial c}{\partial t} = D\left(\frac{\partial^2}{\partial r^2} + 
\frac{1-2q}{r}\,\frac{\partial}{\partial r}\right)c + J\delta({\bf r})\,\theta(t)
\end{equation}
where we have added the source term. The comparison of \eqref{conv_diff:2d} with \eqref{RD:c2d}
explains the assertion \eqref{dim} that the effective dimension is equal to $d=2-2q$. The radial symmetry and the $r^{-1}$ form of the velocity field are of course essential for the above reduction of convection-diffusion to pure diffusion \cite{why_2d}.

Using the procedure detailed in Sect.~\ref{Stefan-problem} we obtain 
\begin{equation}
\label{c2d:conv_diff}
c'' + \big(\tfrac{1-2q}{\xi}+2\alpha \xi\big)\,c' = 0
\end{equation}
instead of \eqref{c2d-scal}, from which
\begin{equation}
\label{conv_diff:sol}
c(\xi)=A\int_\xi^1 d\eta\,\eta^{2q-1}\,e^{-\alpha\eta^2}
\end{equation}
The amplitude $A$ and the constant $\alpha$ are related through the same relation 
\eqref{Aa}. Using again the conservation law \eqref{conserv_law:2d} in conjunction with \eqref{radius:2d} and \eqref{conv_diff:sol} we obtain 
\begin{equation}
\label{conv_diff:aJD}
\alpha + A\alpha\int_0^1 d\eta\,\eta^{1+2q} e^{-\alpha\eta^2} = \Phi
\end{equation}

When $q=0$ we recover our previous results. Here are two more examples.

\subsubsection{Effectively zero-dimensional case}

When $q=1$, or equivalently $d=0$, the integral on the left-hand side of equation \eqref{conv_diff:aJD} is elementary and \eqref{conv_diff:aJD} simplifies to 
\begin{equation}
\label{0d:aJD}
e^{\alpha} - 1 = \Phi
\end{equation}
This effectively zero-dimensional case is easier than the original two-dimensional case as the density \eqref{conv_diff:sol} also significantly simplifies:
\begin{equation}
\label{0d:sol}
c(\xi) = (1+\Phi)^{1-\xi^2} - 1
\end{equation}

\subsubsection{Effective $d=-2$ dimension}

When $q=2$, or equivalently $d=-2$, equation \eqref{conv_diff:aJD} simplifies to 
\begin{equation}
\label{-2d:aJD}
2\,\frac{e^{\alpha} - 1 -\alpha}{\alpha}= \Phi
\end{equation}
while the density \eqref{conv_diff:sol} becomes
\begin{equation}
\label{-2d:sol}
c(\xi) = \xi^2\,e^{\alpha(1-\xi^2)} - 1 +\frac{e^{\alpha(1-\xi^2)} - 1}{\alpha}
\end{equation}

\subsection{Survival Probability}

Assuming the validity of the scaling ansatz \eqref{scaling_P} we obtain
an ordinary differential equation 
\begin{equation}
\label{P:conv_diff}
\mathcal{P}'' + \big(\tfrac{1-2q}{\xi}+2\alpha \xi\big)\,\mathcal{P}' +4\alpha(1+\beta)\mathcal{P}= 0
\end{equation}
for the scaled probability distribution. Transforming again the variable $\xi$ to $\zeta$ which is defined by Eq.~\eqref{zeta:def}, we recast \eqref{P:conv_diff} into 
\begin{equation}
\label{Pzeta:conv_diff}
\zeta\,\frac{d^2 \mathcal{P}}{d\zeta^2} + (1-q-\zeta)\,\frac{d \mathcal{P}}{d\zeta} 
-(1+\beta)\mathcal{P}= 0
\end{equation}
This hypergeometric equation has two linearly independent solutions, 
\begin{equation*}
F(1+\beta, 1-q;  -\alpha\xi^2)~~\& ~~
\xi^{2q} F(1+\beta + q, 1+q;  -\alpha\xi^2)
\end{equation*}
and for $q>0$ the latter solution is selected: 
\begin{equation}
\label{P:conv_diff_sol}
\mathcal{P} = C\,\xi^{2q} F(1+\beta + q, 1+q;  -\alpha\xi^2)
\end{equation}
The boundary condition \eqref{P:BC} yields equation 
\begin{equation}
\label{bqa}
F(1+\beta + q, 1+q;  -\alpha)=0
\end{equation}
which determines the persistence exponent $\beta=\beta(q,\alpha)$. The normalization requirement \eqref{P:condition} allows one to fix the amplitude $C$ in \eqref{P:conv_diff_sol} to yield
\begin{equation}
\label{P:CD_sol}
 \mathcal{P} = (2+2q)\,\xi^{2q}\,\frac{F(1+\beta + q, 1+q;  -\alpha\xi^2)}{F(1+\beta + q, 2+q;  -\alpha)}
\end{equation}

More explicit results can be obtained in the small flux limit. In this situation $\alpha\simeq \Phi$ as it follows from \eqref{conv_diff:aJD}, and when $\alpha\to 0$ the persistence exponent diverges $\beta\to\infty$. In the limit
\begin{equation*}
\alpha\to 0, \quad \beta\to\infty, \quad x=\alpha\beta=\text{finite}
\end{equation*}
the same computation as in \eqref{Bessel_0} allows one to express the confluent hypergeometric function via the Bessel function
\begin{equation}
\label{hyper_Bessel}
F(\beta, 1+q; -\alpha)\simeq \Gamma(q +1)\,\frac{J_q(2\sqrt{x})}{x^{q/2}}
\end{equation}
Comparing \eqref{bqa} and \eqref{hyper_Bessel} we get $2\sqrt{\alpha\beta}=x_{1,q}$, where $x_{1,q}$ is a root of the Bessel function of order $q$, that is $J_q(x_{1,q})=0$; there are infinitely many such roots and $x_{1,q}$ is actually the smallest positive root. Thus in the $\alpha\to 0$ the persistence exponent diverges as
\begin{equation}
\beta=\frac{(x_{1,q})^2}{4\alpha} 
\end{equation}
Using the asymptotic form \eqref{hyper_Bessel} we find that the probability distribution \eqref{P:CD_sol} simplifies to 
\begin{equation}
\label{P:Bessel}
\mathcal{P} = \frac{x_{1,q}\,\xi^q J_q(x_{1,q}\,\xi)}{J_{1+q}(x_{1,q})}
\end{equation}
The physical requirement that the probability distribution must be non-negative explains why the proper root $x_{1,q}$ is the smallest positive root of the Bessel function. 

The probability distribution \eqref{P:CD_sol} also simplifies when $\beta$ is integer. For instance, when $\beta=1$ we get $\alpha=1+q$. In this case
\begin{equation*}
S\sim t^{-1}, \quad \mathcal{P} = 2(1+q)\xi^{2q}(1-\xi^2)\,e^{(1+q)(1-\xi^2)}
\end{equation*}


As an example, consider effectively zero-dimensional case. When $q=2$, or equivalently $d=0$, Eqs.~\eqref{P:conv_diff_sol} and \eqref{bqa} become
\begin{equation}
\label{P:d0}
\mathcal{P} = C\,\xi^2 F(2+\beta, 2;  -\alpha\xi^2)
\end{equation}
and
\begin{equation}
\label{ba:d0}
F(2+\beta, 2;  -\alpha) = 0
\end{equation}
When $\alpha\to 0$, the persistence exponent diverges $\beta\to\infty$. The same computation as in \eqref{Bessel_0} shows that in this limit the confluent hypergeometric function that appears in \eqref{ba:d0} can be expressed via the Bessel function of the first order
\begin{equation*}
F(2+\beta, 2;  -\alpha) \simeq \frac{J_1(2\sqrt{\alpha\beta})}{\sqrt{\alpha\beta}}
\end{equation*}
Therefore  in the $\alpha\to 0$ limit 
\begin{equation}
\beta=\frac{(x_1/2)^2}{\alpha} =  46.12477111\,\frac{D}{J}
\end{equation}
where $x_1=3.83170597$ is the smallest positive root of the Bessel function of the first order and in deriving the second equality we have used the asymptotic $\Phi\simeq \alpha$. In this limit the probability distribution \eqref{P:d0} also simplifies and acquires a neat form 
\begin{equation}
\label{P:limit_2}
\mathcal{P} = \frac{x_1 \xi J_1(x_1\xi)}{J_2(x_1)}
\end{equation}

\section{Summary}
\label{summary}

We have studied the behavior of the two-variable survival probability $S(t_0,t)$. We have mostly focused on the dependence on the observation time $t$ in the limit when it greatly exceeds the release time, $t\gg t_0$. We have demonstrated that in two dimensions the survival probability exhibits an algebraic decay \eqref{survival:2d}. We have also shown that the persistence exponent is related to the dimensionless flux via a root of the confluent hypergeometric function. Similar results continue to hold when mobile $A$ atoms are injected with a fluid, so their transport is determined both by convection and diffusion. 

In three dimensions, we have shown that the survival probability exhibits a stretched exponential decay \eqref{survival:3d}. Furthermore, we have computed the asymptotic behavior of the two-variable survival probability for (almost) arbitrary release and observation times. 

We haven't yet derived the announced result \eqref{surv_asymp} in one dimension. To fill this gap we recall the well-known result (see e.g. Refs.~\cite{Lar,Stefan}) that in one dimension the droplet grows a little bit faster than diffusively, namely there is a logarithmic correction $R=[2Dt\,\ln(J^2 t/D)]^{1/2}$ to the diffusive growth. Using e.g. the approach of Ref.~\cite{RK} one finds that the survival probability remains finite in the $t\to\infty$ limit; in other words, there is a finite chance that an $A$ atom will never meet a $B$ atom, that is, it will forever remain inside the growing droplet.


\begin{thebibliography}{99}
\bibitem{why_2d}
      In arbitrary spatial dimension $d_\text{phys}$ 
      the convection-diffusion problem with radial velocity field $V=\frac{Q}{2\pi r}$ 
      reduces to pure 
      diffusion in the space of effective dimension $d=d_\text{phys}-2q$. However,
      the velocity field $V=\frac{Q}{2\pi r}$ makes sense only in two dimensions where it is an exact solution
      of hydrodynamic equations (corresponding to the incompressible fluid).
\end{thebibliography}
\end{document}